# Response of G-NUMEN LaBr$_3$(Ce) detectors to high counting rates

Elisa Maria Gandolfo[1,2], José Roberto Brandao Oliveira[3], Luigi Campajola[1,2], Dimitra Pierroutsakou[2], Alfonso Boiano[2], Clementina Agodi[4], Francesco Cappuzzello[4], Diana Carbone[4], Manuela Cavallaro[4], Irene Ciraldo[4], Daniela Calvo[5], Franck Delaunay[4,6,7], Canel Eke[8], Fabio Longhitano[9], Nilberto Medina[10], Mauricio Moralles[10], Diego Sartirana[5], Vijay Sharma[11], Alessandro Spatafora[4], Dennis Toufen[12] and Paolo Finocchiaro[4,*]
for the NUMEN collaboration.

[1] Department of Physics "Ettore Pancini" - University of Napoli Federico II, Napoli, Italy;
[2] INFN – Sezione di Napoli, Napoli, Italy;
[3] Instituto de Física da Universidade de São Paulo, São Paulo, Brazil;
[4] INFN – Laboratori Nazionali del Sud, Catania, Italy;
[5] INFN – Sezione di Torino, Torino, Italy;
[6] Department of Physics "Ettore Majorana", University of Catania, Catania, Italy;
[7] LPC Caen UMR 6534, Université de Caen Normandie, ENSICAEN, CNRS/IN2P3, F-14000 Caen, France;
[8] Department of Mathematics and Science Education, Faculty of Education, Akdeniz University, Antalya, Turkey;
[9] INFN – Sezione di Catania, Catania, Italy
[10] Instituto de Pesquisas Energéticas e Nucleares, IPEN/CNEN, São Paulo, SP, Brazil
[11] Department of Radiation Oncology, University of Maryland, Baltimore, MD, USA;
[12] Federal Institute of Education, Science and Technology of São Paulo, 07115-000 Guarulhos, São Paulo, Brazil
* Correspondence: finocchiaro@lns.infn.it

**Abstract:** The G-NUMEN array is the future gamma spectrometer of the NUMEN experiment (Nuclear Matrix Element for the Neutrinoless double beta decay), to be installed around the object point of the MAGNEX magnetic spectrometer at the INFN-LNS laboratory.. This project aims at exploring Double Charge Exchange (DCE) reactions in order to obtain crucial information about the neutrinoless double beta decay (0νββ). The primary objective of the G-NUMEN array is to detect the gamma rays emitted from the de-excitation of the excited states populated via DCE reactions   with good energy resolution and detection efficiency, amidst a background composed of transitions from competingreaction channels with far higher cross-sections. To achieve this, the G-NUMEN signalswill be processed in coincidence with those generated by the detection of the reaction ejectiles in the MAGNEX Focal Plane Detector(FPD). Under the expected experimental conditions, G-NUMEN detectors will operate at high counting rates, of the order of hundreds of kHz per detector, while maintaining excellent energy and timing resolutions. The complete array will consist of over 100 LaBr$_3$(Ce) scintillators. Initial tests have been conducted on the first detectors of the array, allowing for the determination of their performance at high rates.

## 1. Introduction

Understanding the nature of neutrinos is one of the major open quests in the physics beyond the Standard Model. Among the various approaches to this topic, Neutrinoless Double Beta Decay (0νββ) is a unique tool to verify if neutrinos are Majorana particles. In principle, the partial half-life of this decay, together with the Nuclear Matrix Elements (NME), should allow for the determination of the effective Majorana neutrino mass. However, current theoretical calculations and experimental attempts to determine NMEs have not yielded conclusive results [1].

The NUMEN project proposes a new approach to this challenge. It aims at obtaining information on the NMEs of the $0\nu\beta\beta$ decay through the study of the heavy-ion induced Double Charge Exchange (DCE) reactions [2-5]. This approach is based on theoretical analogies between the $0\nu\beta\beta$ decay and the DCE reactions, which have been extensively investigated in previous works [6-8] and which strongly support the correlation between these two reactions and the development of DCE-constrained theories for the NME of the $0\nu\beta\beta$ decay.

The NUMEN project focuses on exploring DCE reactions for both isospin-lowering ($\tau^-\tau^-$) and isospin-rising ($\tau^+\tau^+$) directions, in analogy with the $\beta^-\beta^-$ and $\beta^+\beta^+$ decays. In a long term perspective, this study involves several $0\nu\beta\beta$ decay candidate isotopes as targets such as $^{76}$Ge, $^{82}$Se, $^{110}$Pd, $^{124}$Sn, $^{116}$Cd, $^{130}$Te, $^{136}$Xe, and heavy-ion beams of $^{18}$O$^{8+}$ for $\beta^+\beta^+$ and $^{20}$Ne$^{10+}$ for $\beta^-\beta^-$, with energies ranging from 15 to 60 AMeV. In the explored collisions, other reactions compete with the DCE channel such as multi-nucleon transfer [9-12], single charge exchange [13,14], elastic and inelastic channels [15-17], besides deep inelastic, fusion-fission, fusion evaporation processes and others, with cross-sections up to several orders of magnitude higher than the DCE. Indeed, the main experimental challenge for NUMEN lies in the few nb cross section expected for the DCE reactions [18] and the consequent need for a high-performance detection apparatus with high sensitivity and resolution and excellent discrimination capabilities. Such an apparatus is currently under development [19].

The detection apparatus comprises the high-acceptance MAGNEX spectrometer, together with its focal plane detector (FPD) [20,21], and the G-NUMEN array consisting of 110 LaBr$_3$(Ce) scintillators placed around the target area (Figure 1) [4,22,23].

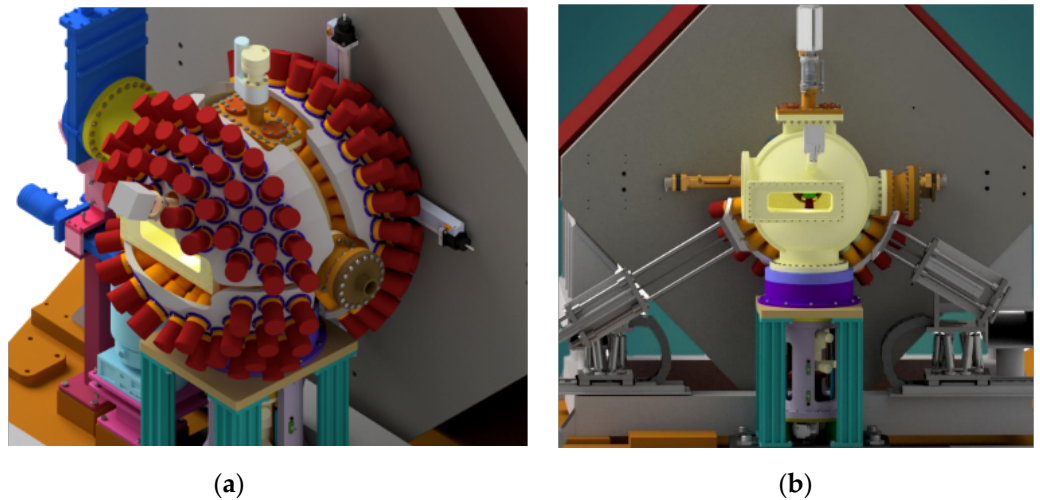

(a)            (b)

Figure 1. Design of the G-NUMEN array. (**a**) The LaBr3:Ce detectors are placed around the scattering chamber, supported by individual mechanical shells. (**b**) The scattering chamber (in yellow) gives a constraint for the minimum distance possible (24 cm) between target and scintillators and for the maximum solid angle coverage.

MAGNEX can detect the DCE events with excellent selectivity [24]. Its energy resolution is normally sufficient for the separation of low-lying energy states involving non-deformed nuclei in experiments with low-energy beams, while the G-NUMEN array will be crucial for all the situations involving high-energy beams and deformed target nuclei for which the resolution of MAGNEX will not be enough to separate nearby states of the residual nuclei under study. In these cases, the DCE states will be identified by time coincidence and anti-coincidence between the MAGNEX Focal Plane charged particle detectors (FPD) and the characteristic gamma-ray transitions detected by G-NUMEN. Indeed, in a typical G-NUMEN experiment an almost continuum gamma-ray spectrum is expected, and identifying the DCE gamma transition lines requires the coincidence with MAGNEX, according to detailed GEANT4 simulations [4].

The main LaBr$_3$(Ce) characteristics of interest for NUMEN, besides its radiation hardness, particularly to fast neutrons, are its high light output, the excellent energy resolution and the fast time response that fit the very strict demands for the G-NUMEN array. The most important demands are a time resolution of a few ns in order to able to distinguish between subsequent beam bunches (occurring every 30-50 ns) and to reject very intense background, and an energy resolution better than 10% at 200 keV, enough to separate the first and second excited states of the nuclei under study [4].

Extensive literature can be found about the crystal characterization and its scintillation properties [25-29]. However, due to the extreme experimental conditions foreseen for the reactions involving G-NUMEN, the simulations show that the G-NUMEN array will be exposed to an intense radiation background of both neutrons and gammas resulting in detection rates up to 300 kHz for each scintillator [4].

This challenging experimental environment requires further characterization of the scintillators performance in order to deeply understand the response of the array under such detection rates, in particular their energy resolution, their linearity and the optimal electronic setup for signal processing. This work focuses on the characterization and testing of the first prototypes of the G-NUMEN scintillators, evaluating their performances under the foreseen detection rate for the NUMEN experiment. Different electronic systems for the signal acquisition have been assessed with the purpose of identifying the most suitable setup for the future experimental conditions.

## 2. Materials and Methods

The material, geometry and dimensions of the G-NUMEN scintillator array have been carefully selected to optimize timing, efficiency, energy resolution and signal-to-noise ratio, while also adapting to various mechanical constraints. Each scintillator consists of a LaBr3(Ce) crystal (ϕ = 38 mm diameter, l = 50 mm length) whose light is collected by an 8 dynode-stage PhotoMultiplier Tube (PMT). The PMT is R6231 model from Hamamatsu while the crystals are produced by Epic-Crystal (China). Figure 2 shows a typical spectrum obtained with a 22Na source which produces gamma rays of 511 and 1275 keV. Also shown is the channel to energy linear calibration plot, produced using the data in Table 1.

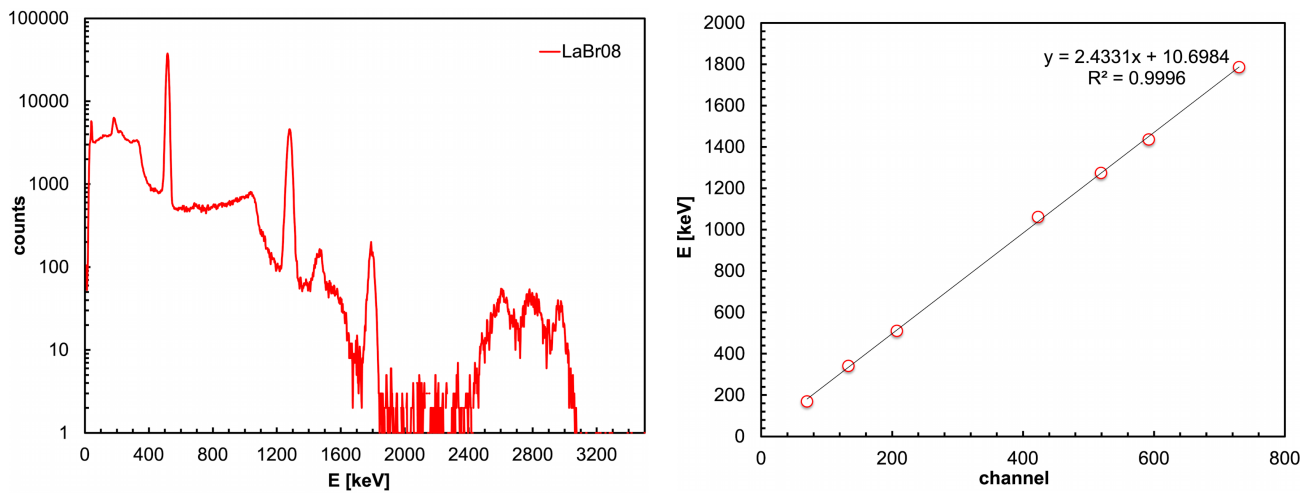

Figure 2. (a) Typical spectrum obtained with one of the LaBr$_3$ detectors exposed to a $^{22}$Na source which produces gamma rays of 511 and 1275 keV. The voltage bias was 750 V. (b) The channel to energy linear calibration plot.

Table 1. Data used for the channel to energy linear calibration of Figure 2.

| Energy [keV] | Channel | note |
|---|---|---|
| 170 | 70 | backscattering of 511 keV gamma |
| 341 | 133 | compton shoulder of 511 keV peak |
| 511 | 207 | annihilation of $^{22}$Na positron |
| 1062 | 423 | compton shoulder of 1275 keV peak |
| 1275 | 519 | $^{22}$Na main gamma peak |
| 1436 | 592 | $^{138}$La --> $^{138}$Ba --> gamma |
| 1786 | 730 | 511+1275 pileup |

Figure 3 shows the prototype of one shell of the gamma array and the typical background spectrum, with prominence of the internal radioactivity of the detectors. To assess the performances of the scintillators, key parameters such as energy resolution, efficiency, gain and linearity have been evaluated. Calibration sources such as $^{22}$Na, $^{60}$Co, $^{137}$Cs, $^{152}$Eu have been used for this purpose, as well as the internal activity peaks of the crystal, at 35.5 keV and 1472 keV [29-30].

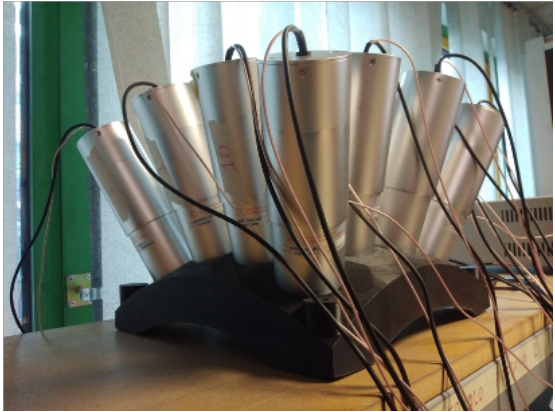
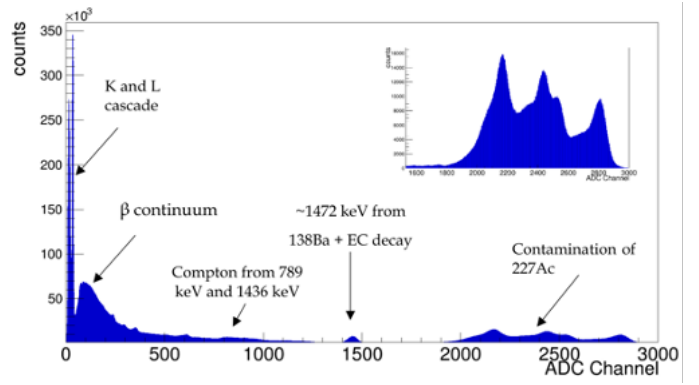

(**a**)                                                              (**b**)

Figure 3. (**a**) The prototype of one shell of the G-NUMEN array. (**b**) The typical background spectrum of the scintillators to be used in the NUMEN experiment.

The characteristics of the scintillators have been tested at increasing detection rates to obtain a global description of the detectors response to counting rates up to the maximum rate foreseen in NUMEN, i.e. ~ 300 kHz. To reach these high rates, different approaches were employed.

An intense $^{137}$Cs radioactive source with activity of 1.5 MBq was used at the INFN-LNS facility, and placed at varying distance from the detectors to vary their counting rate. The intense source was used in addition to calibration sources, whose peaks were analyzed in terms of energy resolution and gain. This allowed for an investigation of the scintillator responses under different levels of counting rate up to 180 kHz. The detector signal pulses were sent to a CAEN VX2745 16 bit 125MS/s digitizer, running a Pulse Shape Discrimination firmware which allows to integrate the pulse charge for an adjustable time range.

The detectors were also exposed to a fusion-evaporation reaction at the ALTO facility of the IJC Laboratory. This reaction induced high detection rates up to 310 kHz and provided additional insights into the scintillator performances. In this case a $^{22}$Na calibration source was used as reference, whose emitted gamma rays were detected over the background produced by the fusion-evaporation reaction. The average energy of the background spectrum during these tests are similar to  typical G-NUMEN DCE experiments, around 400 keV. Indeed, the experiments presented in this work aimed at simulating realistic experimental conditions and evaluating the behavior and response of the scintillators at the anticipated detection rates expected for the NUMEN project.

To explore the compatibility of different electronic configurations with the foreseen experimental environment, the tests were conducted at different supply voltages (HV) using both passive resistor-type (Hamamatsu E1198-26) and active-type (EPIC) voltage dividers. Indeed, the detector performances can be affected by the detection rate as well as by the voltage supply and the electronic configuration (i.e. voltage divider type).

The scintillation yield of the LaBr$_3$(Ce) crystals of the various detectors is expected to be very similar, but the quantum efficiency and gain of the photomultipliers, even those of the same model and at the same supply voltage, can be quite different for each individual device. The output pulse charge per gamma-ray energy was observed to vary significantly, at the same applied voltages, among the 15 prototype detectors which were tested. In addition, the performance of the PMT bases used is impacted by the average currents drawn from the PMT. Therefore, it is crucial to perform experimental tests to evaluate the response of the various detectors by varying the voltage bias, the detection rate and the electronic configurations.

## 3. Results

### 3.1. Energy resolution and photopeak efficiency at low rates with different base configurations

The energy resolution and photopeak efficiency of two of the detectors are shown in Figure 4. The data were collected at fixed voltage (HV = -1000 V) and low detection rate (< 10 kHz). In this condition, these characteristics depend mainly on the crystal, therefore no changes can be noticed between the results for the resistor-type and active-type voltage divider (aka base).

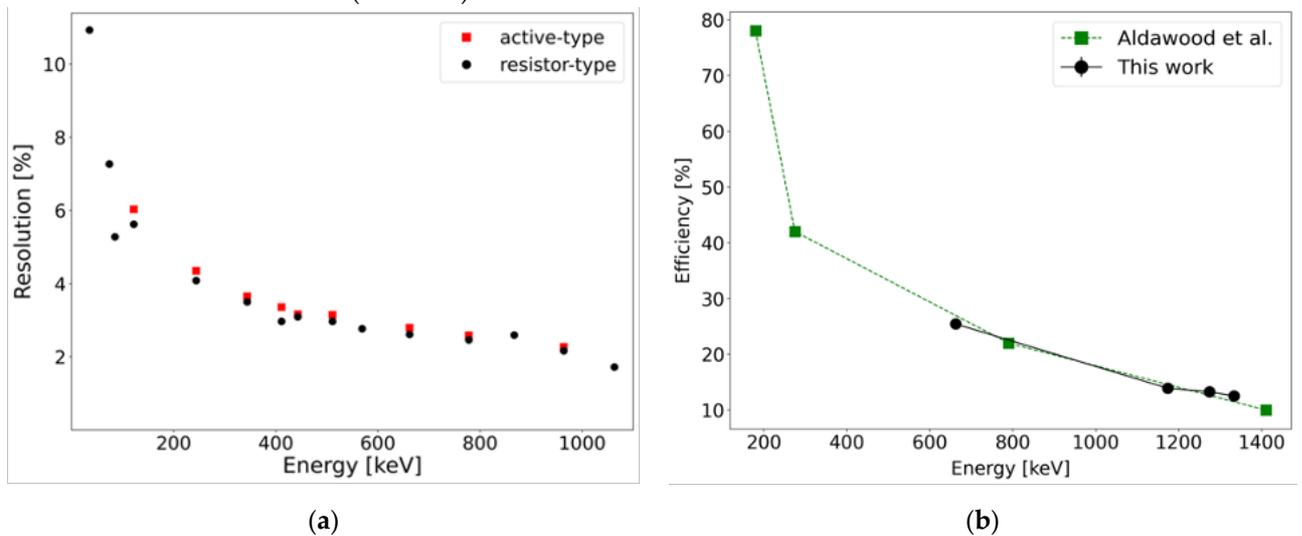

(**a**)    (**b**)

Figure 4. (**a**) Energy resolution characterization of the detectors (at low counting rates) in the two configurations with resistor-type (black) and active-type (red) voltage dividers using $^{22}$Na, $^{60}$Co,$^{137}$Cs, $^{152}$Eu, calibration sources and the self-activity peaks. (**b**) The measured intrinsic photopeak efficiency, in agreement with literature data [25,26].

### 3.2. Gain and non-linearity of the detector response

The charge conversion gain $g$ can be defined as the ratio of the charge and the energy of a given signal ($g = Q/E$). This gain (or integral gain) was measured as a function of the detection rate using an intense $^{137}$Cs source (1.5 MBq). The maximum rate reached in this test was $R_{MAX}$ = 180 kHz. The average output anode current can be obtained from the average pulse charge times the count rate ($I_a = Q_{av}R$). An increase of the detection rate represents an increase of the anode current $I_a$. Figure 5a shows the relative variation of this gain with the anode current at different energies for both the resistor-type and active-type detectors.

It can be noticed that, although both detectors were exposed to the same maximum rate, the maximum anode current is different because the amplification of the two photomultipliers is different already at a low rate. The detector with the resistor-type voltage divider shows a larger relative gain variation with rate, and a substantial difference in the trend of the gain variation at high energy (2.6 MeV) with respect to the one at lower energy (662 keV): this behavior is indicative of an increasing non linearity of the response with energy as the rate is increased. As expected from literature, the response of the detector with the active-type voltage divider is more stable with respect to the resistor-type at high rates.

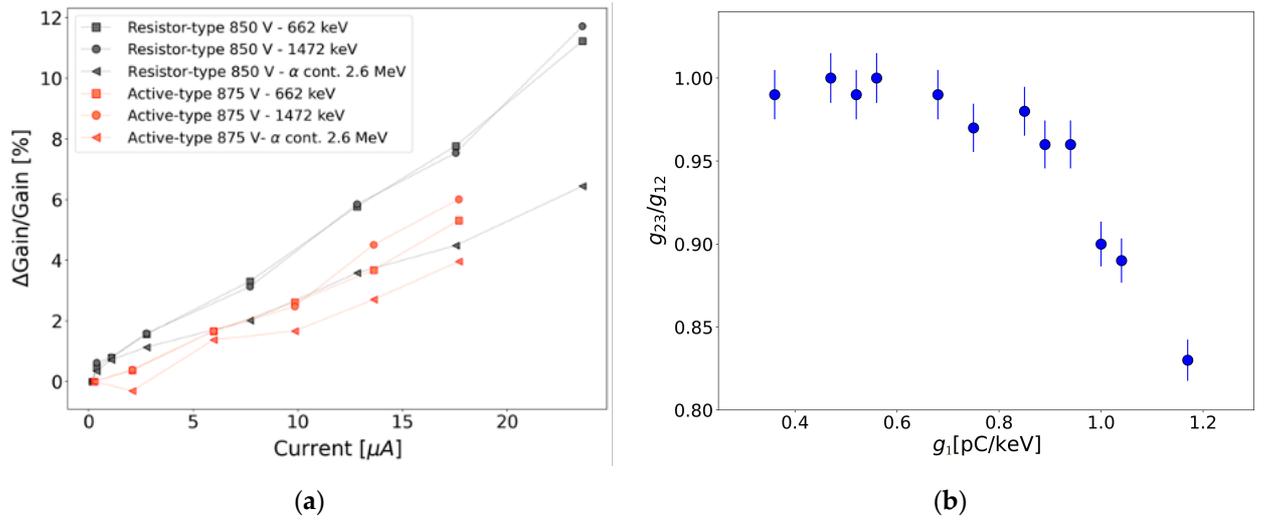

(a)    (b)

Figure 5. (a) Relative variation of the gain with the detection rate (i.e. anode current) for the scintillator with a resistor-type (black) and an active-type (red) voltage divider. The different data points represent the energy peaks of the $^{137}$Cs (square), the intrinsic activity peak at 1472 keV (circle) and the last peak of the intrinsic activity related to the α contamination of the crystal (triangle). (b) Differential gain ratio as a function of integral gain for different detectors at low rates. The PMT supply voltage for all detectors was -937.5 V. Each data-point corresponds to a different detector.

It can be useful to introduce the differential gain $g_{ij}$ which allows for the comparison of the conversion gain at different energies and it can be defined as:

$$g_{ij} = \frac{Q_j - Q_i}{E_j - E_i} \qquad (1)$$

where $Q_j$ is the integrated pulse charge corresponding to the j-th peak centroid of energy $E_j$. The differential gain is more sensitive than the integral gain ($g_j = Q_j/E_j$) to the non-linearity of the conversion. In particular, the deviation from unity of the differential gain ratio $g_{ij}/g_{ni}$, evaluated for a set of peaks in a given energy range, reflects the non-linearity of the conversion in that energy range.

Figure 5b shows the differential gain ratio for several detectors with the active-type base at the energies corresponding to the 1173.2 keV (peak 1), 1332.5 keV (peak 2) transitions from a $^{60}$Co source and 2505.7 keV (peak 3), from their sum peak at close distance geometry, all with the same voltage applied to the PMT (-937.5 V). Each data-point corresponds to a different detector. The variation in $g_1$ spans a factor larger than 3 illustrating a wide spread of PMT multiplication factor or gain. The drop of $g_{ij}/g_{ni}$ observed at large $g_1$ indicates the emergence of a non linear behavior due to space-charge effects as the pulse charge increases. The PMT differential gain $g_{12}$ as a function of the applied voltage can be seen on Figure 6a for two different detectors. Note that at this gamma energy range, both the differential and integral gains are very similar. This dependence can be expressed as:

$$g = p_0 V^{p_1} \qquad (2)$$

where $p_0$ and $p_1$ are the adjustable parameters of a fit to the $g$ versus $V$ data of a specific detector, such as those of Figure 6b.

### 3.3. Energy resolution at different PMT voltages

Eq.2 allows to express the energy resolution of the detector as a function of the gain instead of the supply voltage, as shown in Figure 6b. The highest gain of each detector is obtained at -1000V. Both the low gain PMT detector (red data points) and the large gain PMT detector (blue data points) present a nice resolution (~2%) around the same gain (0.2 pC/keV), rather than at the same voltage. Figure 7 shows a similar plot for different gamma-ray energies, illustrating that roughly the same gain (0.2-0.3 pC/keV) is required for good resolution at all energies.

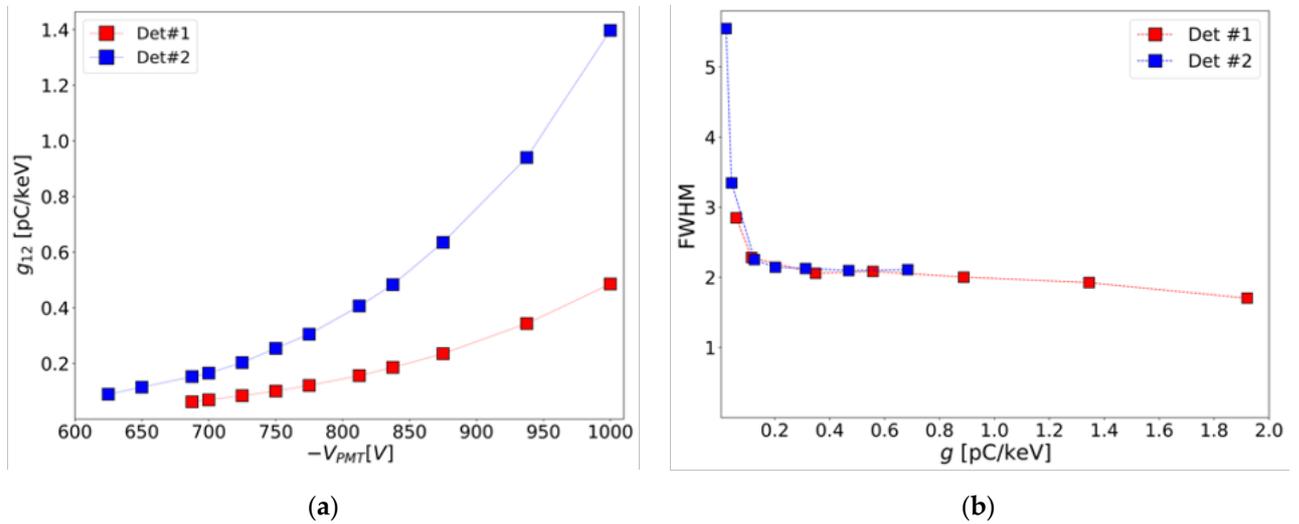

(**a**)                                          (**b**)

Figure 6. (**a**) Differential gain (using the 511 keV and 662 keV gamma peaks) of two active-type detectors as a function of the supply voltage. (**b**) Energy resolution of the 1332 keV peak as a function of the integral gain (varied by application of different PMT voltages) for two active-type detectors with very different PMT gain at the same voltage.

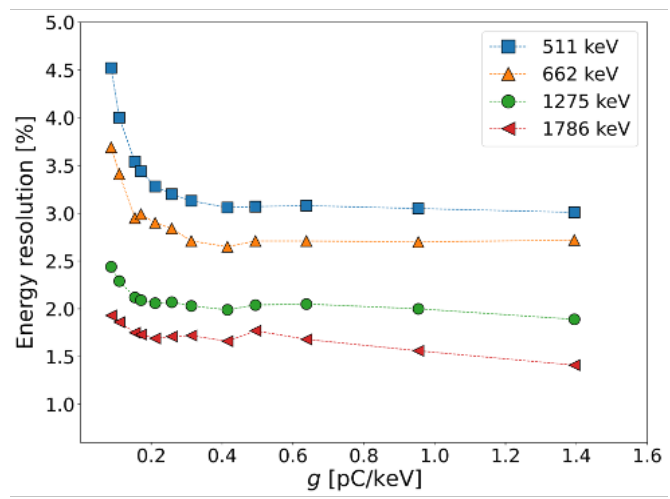

Figure 7. Energy resolution as a function of gain for various gamma-ray energy lines.

*3.4. Limitation in current*

The gain variation of both resistor-type and active-type detectors have been measured in experimental conditions similar to those expected in NUMEN, reaching high detection rate corresponding to a maximum of 180 kHz for the active-type detector and 310 kHz for the passive-type detector. Figure 8a shows the relative gain variation for the two detector types under such counting rates; the latter have been expressed in terms of anode current to be able to compare the different base performances rather independently of the respective PMT gains. The data points in red and black correspond, respectively, to the response of the active-type and the resistor-type base detector to the intense $^{137}$Cs source (max. 180 kHz); the blue points represent the response of the resistor-type detector to the high rate produced in the fusion-evaporation reaction (max. 310 kHz). It can be noticed that the active-type detector shows an abrupt change in gain variation at a rate of ~ 110 kHz while the response of the resistor-type detector is smoother. The results shown in Figure 8a are representative of the limitation in counting rate which is more restrictive for the active-type detector than for the resistor-type detector. However, the passive base presents stronger dependence of the gain with the rate than the active one within the latter operational range.

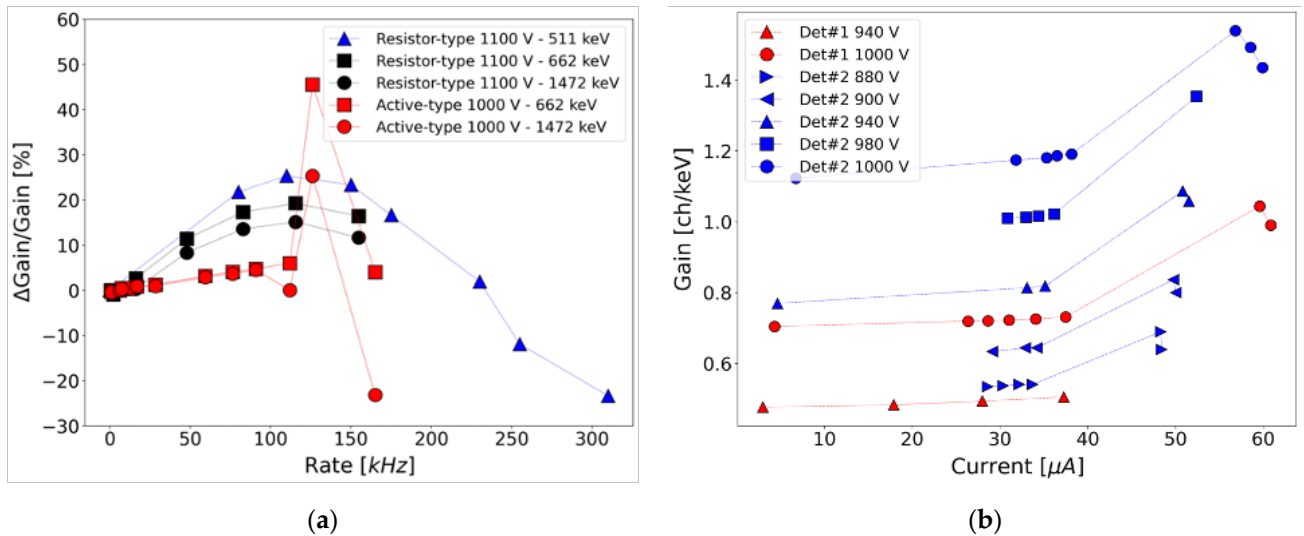

(**a**) (**b**)

Figure 8. (**a**) Variation of the gain with the detection rate at different energies for the resistor-type (black and blue) and the active-type (red) voltage divider. The blue point shows the response of the detector to rate up to 310 kHz obtained with a fusion-evaporation reaction. The black and red point refers to the test performed with an intense $^{137}$Cs source for which the maximum detection rate achieved was 180 kHz. (**b**) Gain at the 662 keV peak as a function of the anode current for two active-type detectors at different voltages.

A more detailed response in gain of the active-type detector is shown in Figure 8b. For all data sets the same maximum count rate of 180 kHz could be reached, but the rates correspond to different anode currents due to the different gains obtained with the individual PMT of each detector.

The data show that the abrupt change in gain occurs for anode currents in the range from 33 to 38 μA for this set of detectors and applied voltages, while the corresponding count rates could be very different. This anode current value represents the limit in current $I_L$ after which a sort of *"breakdown"* of the active base performance is observed.

Figure 9 shows the waveform of a typical detector signal before and after the limiting current is reached: a clear degradation of the signal can be noticed, with a significant increase of rise and fall times.

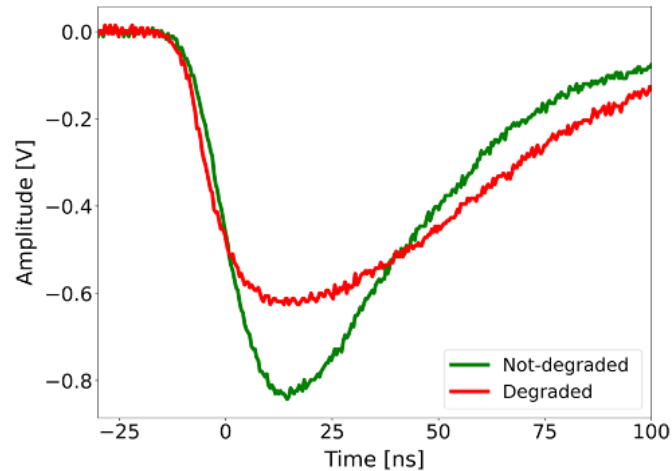

Figure 9. Sample detector signals before (green) and after (red) reaching the limiting current.

*3.5. Systematic measurements of an individual PMT with the active base and laser pulses*

A laser diode, driven by a CAEN DT5810B Dual Fast Digital Detector Emulator, was used to send light directly to the PMT by means of an optical fiber installed sideways with respect to the crystal. This made possible to emulate the detection of signals corresponding to several gamma-ray energies and count rates. The pulse shapes were tuned to be similar to the actual pulses produced by gamma ray detection. The integrated charge was scaled to gamma ray energy, at specific PMT supply voltages, by calibration with data taken with radioactive sources. At low count rates, low voltage bias and gamma-ray energies, this conversion factor is independent of the gamma-ray energy. The laser diode signals were calibrated by means of the simultaneous acquisition of the signals of a $^{60}$Co source and of the LaBr$_3$(Ce) internal radioactivity. All the peaks in the spectrum (internal radioactivity, $^{60}$Co source and laser peak) were analyzed in terms of energy resolution and gain.

The detector performance at high rates was determined with a light pulse corresponding to about 430 keV, that is similar to the average energy from the measurements with a gamma source and to the expected average energy from a typical NUMEN experiment. The plot in Figure 10 shows the limiting current of the PMT with the active base used for the measurements with the laser, as compared with data taken with gamma rays on 3 other detectors.

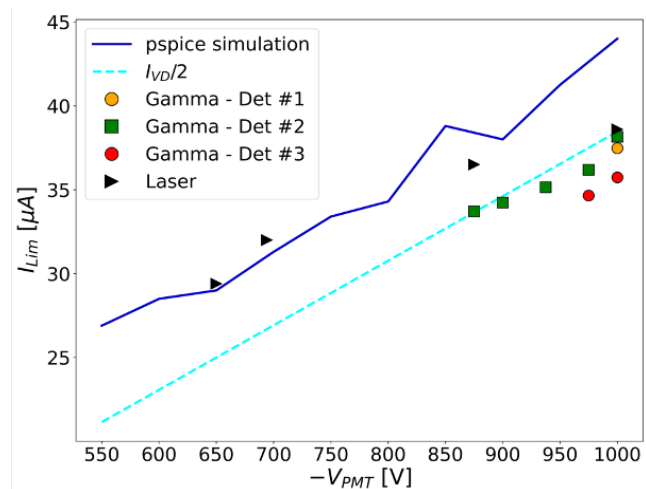

Figure 10. Limiting current versus applied voltage for 4 different detectors (triangles: laser data, squares and circles: gamma data for three other detectors). Solid line: simulations, Dashed line: voltage divider current of the active base divided by two (I$_{VD}$/2, see text).

It can be seen that the limiting current increases with the applied PMT voltage, while unexpectedly similar values are obtained for different detectors in spite of the different amplification factors of the various individual photomultipliers and conversion gains at different voltages (see also Figure 8b). The green data points correspond to a conversion gain spanning more than a factor two in the 875-1000V range, while the green and the yellow data points at 1000 V correspond to detectors with about 1.6 factor in conversion gain. The laser measurements also show that the energy resolution at various energies is constant as a function of the rate up to the limiting anode current.

**4. Simplified modeling of the electronics**

*4.1. The active base electronic schematics*

In the active base circuit used to distribute the voltage to the PMT dynodes, shown in Figure 11, only the potential of the last dynode (DY8) is stabilized by a MOSFET transistor (2SK2168). Figure 10 presents the relevant part of the schematics. The current drawn from the last dynode ($I_8$) added to the one arriving from the MOSFET ($I_Q$) determines the potential drop through the 2MΩ (R3) resistor. As the DY8 current increases, the impedance of the MOSFET also increases reducing $I_Q$, so that the sum ( $I_8+I_Q = I_{VD}/2$) is constant, keeping the potential at the anode ($V_8$) stabilized at only a few volts difference from the Gate (G) potential, controlled by a voltage divider formed by two 2MΩ resistors (R1 and R2). The sum of the currents through R2 and R3 continues to the set of 1MΩ resistors in series that form the rest of the (passive) voltage divider which biases the other dynodes. When $I_Q$ reaches zero (the MOSFET is in cutoff), further increases of $I_8$ cannot be compensated any more and $V_8$ ceases to be stabilized (the DY8 to P potential difference decreases, while the acceleration potential of all previous amplification stages increases, sharply increasing the gain). This happens at $I_8 = I_{VD}/2$ and corresponds to the limiting anode (P) current $I_L$, which should be somewhat larger than $I_8$ due to the current delivered to the last dynode from the previous PMT stage. If the PMT amplification is very large one could assume that $I_L$ is approximately equal to $I_8$. In reality $I_L = I_8 + I_L/G_{DY8}$ (where $G_{DY8}$ is the gain of the last dynode) and should be expected to be 30-40% larger, typically.

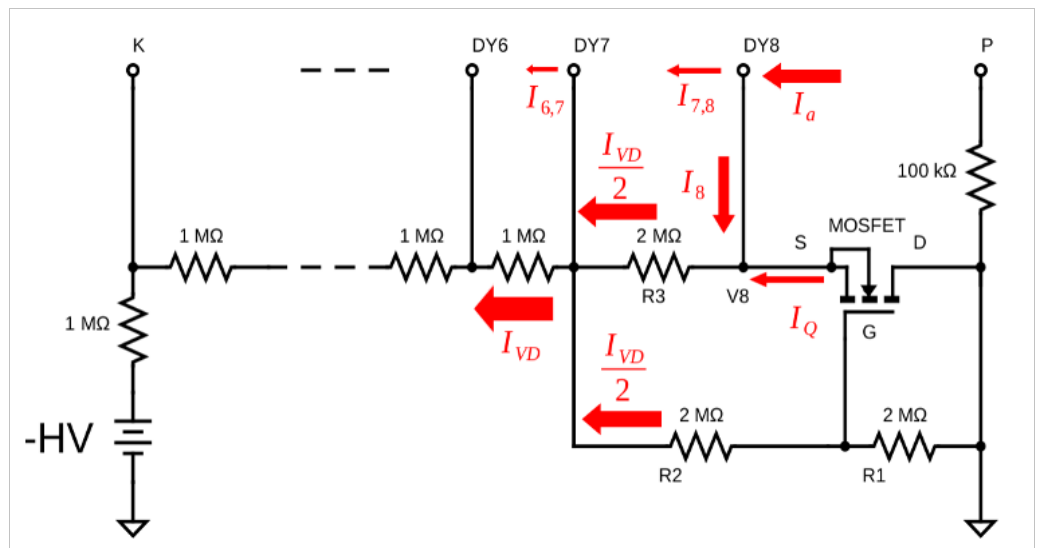

Figure 11. Simplified schematics of the active base highlighting the active stage for stabilization of the last dynode (DY8) potential (V8).

For a given supply voltage, this limit in current corresponds to a limit in the detection rate at which the detector can operate $R_L$. It is possible to calculate the approximate model limit in detection rate as a function of the supply voltage. In fact, $I_L$ can be expressed both in terms of the PMT voltage $V_{PMT}$ and of the detection rate limit $R_L$:

$$I_L \approx \frac{I_{VD}}{2} = \frac{V_{PMT}}{2R_{VD}} \qquad (3)$$

$$I_L = R_L Q_{av} = R_L g E_{av} = R_L p_0 (V_{PMT})^{p_1} E_{av} \qquad (4)$$

where $I_{VD}$ and $R_{VD}$ are respectively the equivalent current and total resistance of the voltage divider, and $Q_{av}$ and $E_{av}$ (*e.g.*, 430 keV) are respectively the average charge and energy of the detector signals. The approximate limiting current $I_L \approx I_{VD}/2$ is represented by the dashed line in Figure 10. By combining the equations 3 and 4 it is possible to obtain the value of the detection rate limit $R_L$ as a function of the conversion gain $g$ and $R_{VD}$:

$$R_L = \frac{g^{((1/p_1)-1)} p_0^{-(1/p_1)}}{2 E_{av} R_{VD}} \qquad (5)$$

Figure 12 shows the calculated value of $R_L$ for the resistor-type and the active-type voltage dividers and at different values of $R_{VD}$ (solid lines). The calculated values approximately agree with the experimental data acquired with the active-type voltage divider used in the tests ($R_{VD}$ = 13 MΩ). It is expected that the limiting current, and thus the limiting rate, should be significantly larger for not too large gain of the last stage, such as $G_{DY8} < 6$, which should be the case. The agreement is, therefore, fortuitous. The MOSFET does not keep the $V_8$ voltage exactly equal to that of its Gate, and the currents are not exactly equal to $I_{VD}/2$ in both $R_2$ and $R_3$. A more detailed description, as in the simulations of the next section, is able to produce more reliable results than the present simple model. Nevertheless, this simple model adequately describes the general trend.

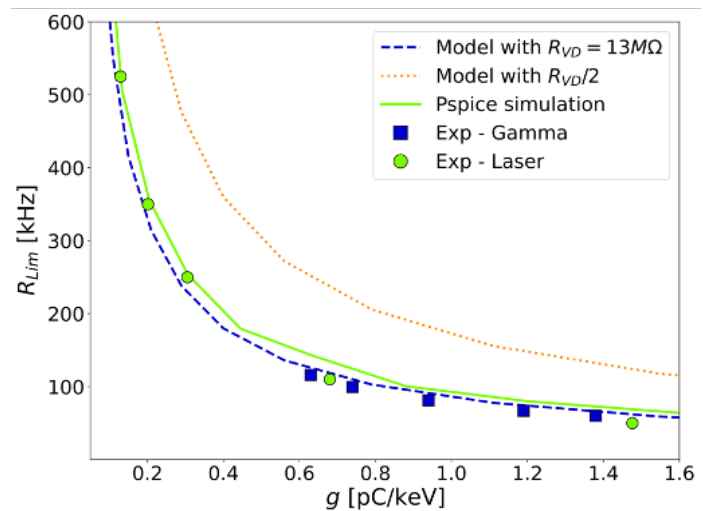

Figure 12. Experimental data (squares and circles), simple model prediction (blue dashed line) and simulated data (green solid line) of the limiting rate as a function of the gain. The orange dotted line is the model prediction by halving the total resistance value of the active-type voltage divider.

*4.2. Simulation of the circuit and PMT response*

The Orcad simulation tool, which utilizes the Pspice model, was employed to simulate the performance of the detectors at high rates. In this simulation, the schematics of the resistor network of the active base and the MOSFET stage, which provides bias to the R6231 Hamamatsu PMT dynodes, together with a model for the dynode stage amplification gain: $G_{DY8} = a(\Delta V)^k$ were considered, where $\Delta V$ is the inter-dynode potential. The parameters $a$ and $k$ were adjusted to reproduce the actual $g$ versus $V$ curve of the PMT used in the laser measurements, and are typical of this type of PMT ($a$ = 0.19906; $k$ = 0.72034). A collection efficiency factor was also taken into account. The solid line of Figure 10 was obtained with this procedure and describes reasonably the laser data. The solid line in Figure 12 represents the result of the simulations for the limiting rate as a function of gain, and is also consistent with the data. The plot in Figure 13 shows the simulated gain variation as a function of the anode current. While the rise in gain within the operational range of the active base observed experimentally is not well reproduced by the simulation, the "breakdown" of the performance occurring at the limiting current is present. Its predicted dependence on the voltage divider values is confirmed, as it is illustrated by the green dashed line.

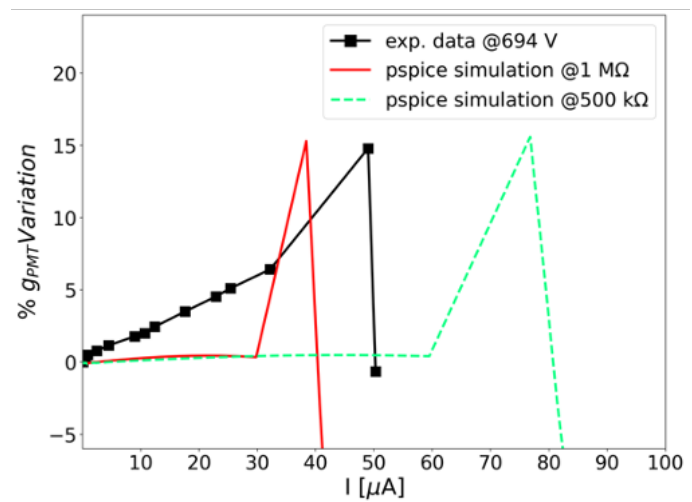

Figure 13. Relative conversion gain as a function of the anode current. Black squares: data points obtained with gamma-ray measurements. Red solid line: simulated gain with value of the currently employed voltage divider resistors (1MΩ). Green dashed line: simulation with halving of the voltage divider resistor values.

## 5. Discussion

The study of the conversion gain and linearity of the detector data shows that there is a variation of the gain with the count rate (and anode current) for both voltage divider types. Indeed the effect is due to a combination of the charge per pulse and the pulse repetition rate. The detector with the resistor-type voltage divider has a larger variation with count rate and exhibits a higher non-linearity in its response to different energies. On the other hand, the configuration with the active-type voltage divider demonstrates a more linear and stable response, making it the preferred choice for the application in the G-NUMEN project.

It is worth noting that the active-type detector still shows a non-negligible dependence of gain and energy resolution on count rate (and anode current), as well as on the voltage supply. However, revision of the base schematics will be studied in order to further reduce the instability and non-linearity of the response. The results obtained from the tests presented in this study establish a connection between the detector response in terms of energy resolution and gain and the changes in the anode current and in the supply voltage.

Figure 6 and Figure 7 allow to extract the minimum gain at which the energy resolution requirements are met. A conversion gain of about 0.35 pC/keV should be sufficient to provide with a good resolution at all energies. This gain value corresponds to a HV in a range of 750-850 V depending on the detector. However, a limitation of the anode current in the active base has been observed. Above a certain current value $I_L$, the detector performance is degraded. The restriction on the anode current also translates into a limitation on the counting rate at which the detector can operate effectively. In section 4.1 it was demonstrated that such a limit, $R_L$, can be estimated from equation 5 by taking into account the HV and the voltage divider total resistance ($R_{VD}$ = 13 MΩ), and that the maximum achievable counting rate for the detector operating with good energy resolution is around 200 kHz. Reducing the voltage divider total resistance could help in achieving the desired count rate, however other alternative designs of the active base are under consideration to achieve also better linearity and stability of the response.

## 6. Conclusions

This works presents the first tests of the prototype of the G-NUMEN array, conducted under experimental conditions similar to those foreseen for the NUMEN experiment. The results demonstrate that the performance of the detectors is strongly influenced by the count rate, as well as by the bias voltage and by the electronic configuration of the voltage divider. The findings of this work indicate that the active-type voltage divider could offer the best electronic configuration for the detector array. However, limitations have been observed in the anode current, with performance degradation occurring after reaching a specific value that depends on the voltage applied to the PMT. The limit in the anode current translates into a restriction on the count rate sustainable by the detector. While operating the PMT at lower voltages could potentially allow for higher count rates, this approach would compromise the energy resolution, which is a crucial requirement for the NUMEN project.

The calculations presented in this study provide valuable insights into determining the optimal electronic configuration of the detector for the future G-NUMEN array. This study has demonstrated that in order to address the challenges encountered at detection rates comparable to those expected in the NUMEN project, improvements to the detector performance can be achieved by modifying the specific circuit of the active voltage divider. These findings contribute to a better understanding of the detector behavior and offer guidance for optimizing its performance under various operational conditions.

**Author Contributions:** Conceptualization, E.M.G., L.C., J.R.B.O. and P.F.; methodology, J.R.B.O., D.P. and P.F.; validation, E.M.G., L.C., J.R.B.O. and P.F.;  formal analysis, E.M.G., J.R.B.O., D.P., A.B. and D.T; investigation, E.M.G., J.R.B.O., L.C., Da.C., Di.C., D.P., A.B., M.C., I.C., F.D., C.E., F.L., N.M., M.M., V.S., D.S., A.S. and P.F.; resources, C.A., F.C., P.F. and F.D.; data curation, E.M.G; writing-original draft preparation, E.M.G. and J.R.B.O.; writing-review and editing, F.D., D.P., A.B., Di.C. and P.F.; visualization E.M.G. and P.F.; supervision, L.C., J.R.B.O. and P.F.; project administration, C.A. and F.C.; funding acquisition, C.A. and F.C. All authors have read and agreed to the published version of the manuscript.

**Funding:** This research was funded by INFN and by the Brazilian funding agencies FAPESP Proc. No. 2019/07767-1, CNPq proc. 316019/2021-6, and INCT-FNA Proc. No. 464898/2014-5.

**Data Availability Statement:** The data presented in this study can be made available on request from the corresponding author.

**Acknowledgments:** The authors wish to thank the radiation protection staff of the LNS facility, as well as the technical staff of the ALTO facility for their precious support and contribution to this work.

**Conflicts of Interest:** The authors declare no conflict of interest.